\documentclass[fleqn,usenatbib]{mnras}

\usepackage{newtxtext,newtxmath}

\usepackage[T1]{fontenc}

\DeclareRobustCommand{\VAN}[3]{#2}
\let\VANthebibliography\thebibliography
\def\thebibliography{\DeclareRobustCommand{\VAN}[3]{##3}\VANthebibliography}

\usepackage{graphicx}	
\usepackage{amsmath}	
\usepackage[dvipsnames]{xcolor}
\usepackage{showyourwork}

\newcommand{\lisa}{{\it LISA}}
\newcommand{\gaia}{{\it Gaia}}

\title[LISA CVs]{Cataclysmic variables are a key population of gravitational wave sources for LISA}

\author[S. Scaringi et al.]{
S. Scaringi$^{1}$\thanks{E-mail: simone.scaringi@durham.ac.uk},
K. Breivik$^{2}$,
T.B. Littenberg$^{3}$,
C. Knigge$^{4}$,
P.J. Groot$^{5,6,7}$ and
M. Veresvarska$^{1}$
\\
$^{1}$Centre for Extragalactic Astronomy, Department of Physics, Durham University, South Road, Durham, DH1 3LE\\
$^{2}$Center for Computational Astrophysics, Flatiron Institute, 162 Fifth Avenue, New York, NY, 10010, USA\\
$^{3}$NASA Marshall Space Flight Center, Huntsville, Alabama 35811, USA\\
$^{4}$School of Physics and Astronomy, University of Southampton, Highfield, Southampton SO17 1BJ, UK\\
$^{5}$Department of Astrophysics/IMAPP, Radboud University, P.O. 9010, 6500 GL, Nijmegen, The Netherlands\\
$^{6}$South African Astronomical Observatory, PO Box 9, Observatory, 7935, Cape Town, South Africa\\
$^{7}$Department of Astronomy, University of Cape Town, Private Bag X3, Rondebosch, 7701, South Africa\\
}

\date{Accepted XXX. Received YYY; in original form ZZZ}

\pubyear{2023}

\begin{document}
\pagerange{?--?}
\maketitle

\begin{abstract}
The gravitational wave (GW) signals from the Galactic population of cataclysmic variables (CVs) have yet to be carefully assessed. Here we estimate these signals and evaluate their significance for \lisa. First, we find that at least three known systems are expected to produce strong enough signals to be individually resolved within the first four years of \lisa's operation. Second, CVs will contribute significantly to the \lisa\ Galactic binary background, limiting the mission's sensitivity in the relevant frequency band. Third, we predict a spike in the unresolved GW background at a frequency corresponding to the CV minimum orbital period. This excess noise may impact the detection of other systems near this characteristic frequency. Fourth, we note that the excess noise spike amplitude and location associated with $P_{\rm{min}}\sim80~\mathrm{min}$ can be used to measure the CV space density and period bounce location with complementary and simple GW biases compared to the biases and selection effects plaguing samples selected from electromagnetic signals. Our results highlight the need to explicitly include the Galactic CV population in the \lisa\ mission planning, both as individual GW sources and generators of background noise, as well as the exciting prospect of characterising the CV population through their GW emission.
\end{abstract}



\section{Introduction}
\label{sec:intro}

The \textit{Laser Interferometer Space Antenna} (\lisa; \citealt{lisa17}) is a space-based gravitational-wave (GW) detector due to launch in 2037. \lisa\ will be sensitive to signals in the frequency range $10^{-5}~\mathrm{Hz} \lesssim f \lesssim 10^{-1}~\mathrm{Hz}$, allowing 
it to detect GWs produced by Galactic compact binary systems among several other source classes including coalescing supermassive black hole binaries \citep[e.g.][]{Klein2016, Bellovary2019}, extreme mass ratio inspirals \citep[e.g.][]{Berti2006, Barack2007}, and cosmological GW backgrounds \citep[][]{Caprini2016, Bartolo2016}.

The binary stars \lisa\ is sensitive to are those with orbital periods ranging from minutes to hours. Among the ultracompact binaries consisting of two stellar remnants, the AM~CVn systems are ultra-compact binaries in which a white dwarf (WD) accretes material from a fully or partially electron-degenerate companion (e.g. another WD). AM~CVn stars have orbital periods in the range $5~\mathrm{min} \lesssim P_{\rm{orb}} \lesssim 65~\mathrm{min}$, near the "sweet-spot" of \lisa's sensitivity curve. Several AM~CVn systems have been identified as so-called \lisa\ verification binaries (\citealt{kupfer18,kupfer23}) as their gravitational wave amplitude, and consequently characteristic strain, are strong enough to be resolvable by \lisa\ within the first four years of operation. AM~CVn stars are part of a larger Galactic population of ultra-compact binaries that includes, for example, detached double-degenerate systems. Some of these will be detectable individually with LISA (e.g. \citealt{burdge20}), but the population as a whole is expected to give rise to a broad-band, unresolved GW signal between $\approx 10^{-4}$ Hz -- $10^{-2}$ Hz, which will limit the mission's sensitivity to other GW sources in this band \citep{nelemans01, ruiter10, nissanke12, korol17, lamberts19, breivik20a, breivik2020b, korol22, thiele2023}.

Given the effort that has been expended on predicting the GW signals produced by AM~CVn stars and detached double degenerates -- both individually and as a population -- it is somewhat surprising that their close cousins, cataclysmic variables (CVs), have not received much attention in this context. CVs are compact binary systems in which a WD accretes material from a hydrogen-rich main-sequence or sub-stellar donor star that fills its Roche lobe. Their orbital periods lie in the range $75~\mathrm{min} \lesssim P_{\rm{orb}} \lesssim 1~\mathrm{day}$ {\bf{($0.02~\mathrm{mHz}\lesssim f_{\mathrm{GW}}\lesssim 0.44~\mathrm{mHz}$)}}, part of which still falls into the \lisa\ band. Perhaps more importantly, the space density of CVs is much higher than that of AM~CVn stars. The best current space-density estimates are $\rho_{CV} = 4.8^{+0.8}_{-0.6} \times 10^{-6}~\mathrm{pc^{-3}}$\citep{pala20} and $\rho_{AM} = 5\pm 3 \times 10^{-7}~\mathrm{pc^{-3}}$\citep{carter13} for CVs and AM~CVn systems respectively. This will cause an elevated background of signals in the relevant frequency range. 

The secular evolution of CVs is driven by angular momentum loss (AML) that initially shrinks the binary orbit, reduces the orbital period and keeps the donor in contact with its Roche lobe. As the donor evolves down the main sequence during this evolution, it is driven slightly out of thermal equilibrium and becomes oversized for its mass. Two AML mechanisms are known to be important in driving CV evolution: magnetic braking (MB) and gravitational radiation (GR). MB refers to AML associated with a magnetised wind from the donor star \citep[e.g.][]{Verbunt1981, Rappaport1983}. Since the donor's spin is synchronised with the binary orbit, this ultimately drains angular momentum from the binary as a whole. By contrast, GR-driven AML is directly associated with the emission of GWs \citep[e.g.][]{Peters1964}. 

Here we briefly outline important characteristics in the observed CV population and how these observations translate to theoretical explanations for CV evolution. We refer the reader to \citet{knigge11} for a more detailed discussion. The orbital period distribution of CVs contains two distinctive features, as illustrated in Figure~\ref{fig:porb}. First, there is an obvious dearth of CVs in the orbital "period gap" between $2~\mathrm{hr} \lesssim P_{\rm{orb}} \lesssim 3~\mathrm{hr}$. Second, there is a "period spike" near the minimum orbital period of $P_{\rm min} \simeq 80~\mathrm{min}$ \citep[e.g.][]{gansicke09}. Both of these features are associated with changes in the structure of the donor star. When a CV reaches the upper edge of the period gap at $\approx 3$ hr, its donor is thought to become fully convective. Such stars cannot support the roughly dipolar magnetic fields found in partially radiative stars, since those fields are anchored in the tachocline (the interface between the radiative and convective zone). As a result, the magnetic field topology of the donor must change at this point, and it is thought that MB-driven AML is also severely reduced. This causes a temporary loss of contact, allowing the donor to shrink back to its MS radius. The system then evolves through the period gap as a detached binary until the AML due to GR (and perhaps also due to residual MB or other mechanisms) has shrunk the orbit sufficiently for contact to be re-established. Mass transfer then resumes at the lower edge of the gap.

Below the period gap, the thermal time-scale of the donor increases significantly faster than the mass-loss time-scale. This causes the donor's mass-radius index, $\mathrm{d}\ln R/\mathrm{d}\ln M$, to evolve from a near-thermal equilibrium value of $\zeta_{th} \simeq 1$ towards the adiabatic value of $\zeta_{ad} = -1/3$, where mass loss occurs at a faster rate than the radius can respond. However, the geometry of a semi-detached binary system implies that the donor must also satisfy an orbital period - density relation. It is easy to show from this that the orbital period must start to increase again once the mass-radius index of the donor reaches $\zeta = 1/3$ \citep[see e.g.][]{knigge11}. In practice, this happens at $P_{\rm{min}} \simeq 80$~min and roughly coincides with the transition of the donor from a stellar to a sub-stellar object which is no longer fusing in its core. The period minimum is expected -- and observed -- to coincide with a pile-up in the CV period distribution, the so-called "period spike" \citep[][]{gansicke09}.

Given their relatively high space density and the distinctive nature of their period distribution, CVs merit a close look as potential GW sources for \lisa. We are aware of only two earlier attempts to do this. First, \cite{HBW90} included CVs as one of the six types of Galactic binary systems for which they estimated GW signals. They found them to be a significant, but sub-dominant population compared to W~UMa stars and "close WD binaries" (they did not explicitly consider semi-detached -- i.e. AM~CVn systems -- as part of the latter group). Second, \cite{MAA00} explicitly considered CVs as a population of interest for \lisa. They predicted the GW signals for systems in a catalogue of CVs available at the time, finding that some of these ought to be detectable by the mission. 

Since these pioneering studies were published, our understanding of the Galactic CV population -- and also our understanding of \lisa\ as a GW detector -- has improved significantly. For example, \gaia\ has provided parallaxes for almost all of the closest CVs. This has made it possible to construct the first (nearly) volume-limited sample of CV. Similarly, our ability to estimate system parameters -- notably component masses -- for CVs in which those parameters have not yet been directly measured is also on considerably firmer ground \citep{knigge06,knigge11,savoury11,carter13}. Finally, the \lisa\ mission concept is now sufficiently advanced -- and supported by well-tested numerical tools -- to allow a careful evaluation of potential GW source populations as viable targets for \lisa \citep[e.g.][]{PhysRevD.107.063004,LEGWORK_joss,LEGWORK_apjs}. The goal of our work here is to leverage these improvements to predict the GW signals produced by the Galactic CV population and to assess if and how they will be detectable by \lisa.

\section{Constructing a mock CV population}\label{sec:mock_CV}

\begin{figure}
	\includegraphics[width=0.5\textwidth]{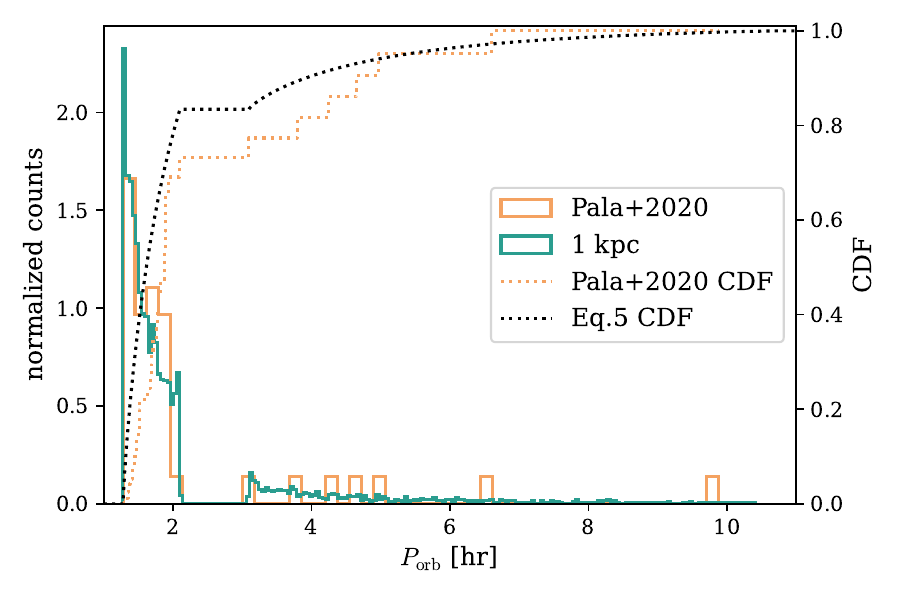}
    \caption{The orbital period distribution of CVs as reported by \citet{pala20} within 150pc (orange line). Also shown is the simulated orbital period distribution within 1kpc (blue line).}
    \label{fig:porb}
    \script{fig1.py}
\end{figure}

The first volume-limited of CVs was selected using parallax measurements provided by \gaia\ was released by \cite{pala20}. The sample consists of $42$ CVs within $150\,\rm{pc}$ and is estimated by them to be $\approx 75\%$ complete. Fig.~\ref{fig:porb} shows the orbital period distribution for these 42 systems in both cumulative (dotted orange curve) and differential forms (thick orange curve).

We generate mock CV samples whose distributions of orbital periods are consistent with that of the intrinsic CV population, but which not restricted to the set of 42 measured values. We require that the mock samples respect the period minimum near $P_{\rm{min}} \simeq 80$~min and the period gap between $2~\mathrm{hr} \lesssim P_{\rm{orb}} \lesssim 3~\mathrm{hr}$ \citep[e.g.][]{gansicke09, knigge06}. Both of these well-known features are present in the \cite{pala20} 150-pc sample. We have approximated the observed cumulative distribution with a simple analytical function:
\begin{equation}
CDF(x) = 
\begin{cases}
      0 & P_{\rm{orb}} < 76.78~\mathrm{min}, \\
      1.98 \left(\frac{x_1 + x_1^{3/4}}{1 + x_1 + x_1^2}\right) & 76.78 \mathrm{min} \leq P_{\rm{orb}} \leq 2.10~\mathrm{hr},\\
      0.834 & 2.10~\mathrm{hr} < P_{\rm{orb}} < 3.09~\mathrm{hr},\\
      0.26 \left(\frac{x_2 + x_2^{3/4}}{1 + x_2 + x_2^2}\right) + 0.834  & 3.09~\mathrm{hr} \leq P_{\rm{orb}} \leq 10.58~\mathrm{hr},\\
      1 & P_{\rm{orb}} > 10.58~\mathrm{hr}
\end{cases}
\label{eq:cdf}
\end{equation}
where $x = \log_{10}({P_{\rm{orb,m}}})$, $x_1 = x - 1.885$, $x_2 = x - 2.269$
and $P_{\rm{orb,m}}$ is the orbital period expressed in minutes. 
We generate mock orbital periods from this via inversion sampling, i.e. by generating $R \sim U(0,1)$, a uniformly distributed random number between zero and unity, and then choosing $P_{\rm{orb}}$ such that $CDF(x) = R$. 

\cite{pala20} have used their 150-pc sample to investigate the intrinsic properties of the Galactic CV population, and, more specifically, to estimate the space density of CVs. Taking into account the missing systems due to completeness, the space density is found to be $\rho_0\simeq4.8 \times 10^{-6}$pc$^{-3}$ in the Solar neighbourhood. This estimate is based on the assumption that the CV population follows an axisymmetric Galactic disc profile of the form
\begin{equation}
\label{eq:disk}
    \rho = \rho_{0} \exp \left(-\frac{|z|}{h}\right)
\end{equation}

\noindent where the CV population scale height $h$ is taken to be 280~pc and $z$ is the distance above the Galactic plane. In reality the scale height $h$ may increase for older CV systems which have shorter orbital periods. However, the local CV population does not allow inference of $h(P_{\rm{orb}})$ due to relatively low numbers of observed CVs. Nevertheless, \cite{pala20} quote different space density values for different scale height assumptions, which are all found to be within a factor of $2$. Here we employ the quoted canonical space density measurement of $\rho_0=4.8 \times 10^{-6}$~pc$^{-3}$ and a scale height of $h=280$~pc.


All but one of the CV systems in the \cite{pala20} sample have known orbital periods. The only exception is \gaia\ J154008.28$-$392917.6, which is thought to be a  WZ~Sge-like system. We therefore set its orbital period to 81 minutes, close to $P_{\rm{min}}$. For simplicity, we adopt an accretor mass of $m_1 = 0.75\, \rm{M_{\odot}}$ for all systems and assign the corresponding donor mass $m_2$ based on the revised evolutionary track from \citet{knigge11} and the observed orbital period. We correct for the incompleteness of the \citet{pala20} 150-pc sample by simulating systems that are distributed to follow the same space distribution. 
The orbital periods of these additional CVs are drawn from the period distribution described by Equation~\ref{eq:cdf}. Component masses are assigned in an identical way as for the observed systems. 

Fig.~\ref{fig:porb} shows the 
orbital period of our $1~\rm{kpc}$ mock population which includes the \citet{pala20} sample in solid teal lines. The cumulative distribution function of the 
$1~\rm{kpc}$ mock population is shown with the black dotted line. 


\section{Gravitational Wave emission from Cataclysmic Variables} \label{sec:gwemission}

Whether a binary is detectable by \lisa\ depends on the GW amplitude of each source, which, when averaged over the \lisa\ constellation orbit, and the system's orbital inclination, depends on the masses of the binary components, the orbital period, and the distance to the source. The GW frequency for a CV (and circular binaries in general) is given by
\begin{equation}
    f = 2/P_{\rm orb},
\end{equation}
with $P_{\rm orb}$ being the binary's orbital period and $f$ being the GW frequency. The GW amplitude is given by
\begin{equation}\label{eqn:amp}
    \mathcal{A} = \frac{2 (G \mathcal{M})^{5/3} }{c^4 d} (\pi f)^{2/3}.
\end{equation}
where $G$ and $c$ are the gravitational constant and speed of light respectively, and $d$ is the distance to the binary. The chirp mass $\mathcal{M}$ is defined as
\begin{equation}\label{eqn:mchirp}
    \mathcal{M} = \frac{(m_1 m_2)^{3/5}}{(m_1 + m_2)^{1/5}}, 
\end{equation}
for component masses $m_1$ and $m_2$.

Since CVs emit at $f_{\rm{GW}} < 1\,\rm{mHz}$, their orbital evolution occurs on long-enough timescales that \lisa\ is unlikely to observe any changes to the GW frequency of any given source. This means that the total signal to noise ratio can be approximated as
\begin{equation}\label{eq:SNR}
       SNR = \frac{\mathcal{A}\sqrt{T_{\rm{obs}}}}{\sqrt{S_n}},
\end{equation}
\noindent where $T_{\rm{obs}}$ is the observation duration, $\mathcal{A}\sqrt{T_{\rm{obs}}}$ is the amplitude spectral density (ASD), and $S_n$ is the power spectral density sensitivity of \lisa. 

The analytic $SNR$ approximation above provides a solid rule of thumb for detectability of a single system, but does not capture the effects of signal confusion from the Galactic population of close, detached DWD binaries, nor the impact of each CV's inclination or the effects of the \lisa\ constellation orbit. In this work we use \texttt{LEGWORK}\footnote{\texttt{LEGWORK} is a Python packaged designed to compute analytic estimates of GW signals for stellar-mass sources with mHz frequencies. Documentation and source code can be accessed at \href{https://legwork.readthedocs.io/}{https://legwork.readthedocs.io/}.} to compute analytic $SNR$ estimates values for both the known and simulated CV populations as discussed below (\citealt{LEGWORK_joss,LEGWORK_apjs}). 


In order to determine how much the $1~\rm{kpc}$ CV population contributes to the Galactic GW foreground for \lisa\, we must include the effects of the Galactic double WD population as well as the \lisa\ constellation orbit. We do this using \texttt{ldasoft} \citep{ldasoft}. For this study, we use the same Galactic population as was used for the LDC2a-v1 data set provided by the \lisa\ Data Challenges (LDCs)~\citep{le_jeune_maude_2022_7132178} which consists of both detached and interacting double WD binaries. We consider two population cases, one which contains only the double WD binaries from the LDC and one which adds our $1~\rm{kpc}$ mock population of CVs before modeling the \lisa\ response to the combined signal of each Galactic binary after four years of operations. Together, these two simulations allow us to capture the effects of considering the $1~\rm{kpc}$ CV population in the Galactic foreground. 

From the simulated data we estimate the noise level by analyzing the \lisa\ time domain interferometry `A' channel. We fit for the variance of the frequency domain residual after removing resolvable binaries from the simulation, determine which of the remaining sources exceed the detection threshold based on the updated estimate of the noise level, and iterate until the detectable population converges. In the \lisa\ literature this residual foreground is often referred to as ``confusion'' noise because it arises from the combined signals from numerous individual sources that can not be individually separated. The method is similar to what is described in \cite{2021PhRvD.104d3019K}.

\section{Results}
\begin{figure}
	\includegraphics[width=0.5\textwidth]{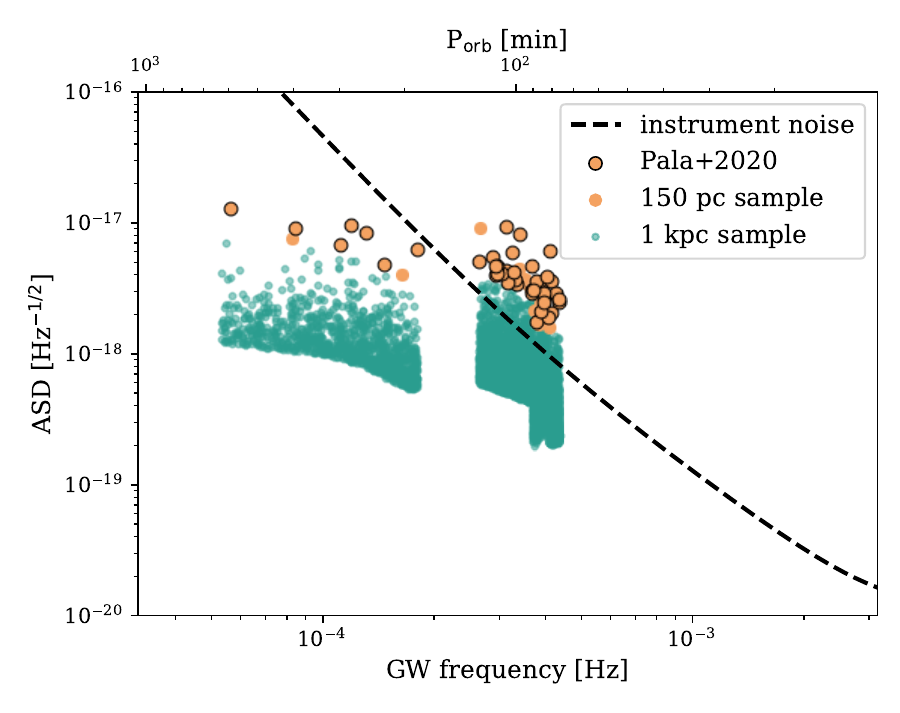}
    \caption{Amplitude Spectral Density (ASD) computed assuming four years of \lisa\ operations as a function of GW frequency. The corresponding \lisa\ instrument noise, without the Galactic foreground, is shown with the dotted black line. The $150~\rm{pc}$ \citet{pala20} sample is shown by the large orange circles, while the remaining simulated systems within $150~\rm{pc}$ are shown in orange. Blue dots show the simulated $1~\rm{kpc}$ CV population using the inferred \citet{pala20} space density.}
    \label{fig:asd}
    \script{fig2.py}
\end{figure}

We first assess the detectability of the populations constructed in Section~\ref{sec:mock_CV} by calculating the ASD assuming a $4~\rm{yr}$ observation duration for each individual CV as shown in Fig.~\ref{fig:asd}. The $150~\rm{pc}$ simulated population is shown in orange filled points with the observed \citet{pala20} systems highlighted with black borders. The $1~\rm{kpc}$ simulated population is shown in teal filled points and the \lisa\ instrument noise curve from \citet{Robson2019}, without the detached DWD confusion noise, is shown by the black dotted line. Although several systems are found to lie close to the conventionally adopted \lisa\ sensitivity limit, 3 systems in particular attain a signal-to-noise detection $>4\sigma$. Unsurprisingly, these are the 3 closest systems, namely WZ Sge, VW Hyi and EX Hya. 

It is also interesting to note that there may be the possibility of detecting other CV systems through their GW emission which have been missed due to electromagnetic detection biases and not included in the current $150~\rm{pc}$ volume-limited sample. In our simulated $150~\rm{pc}$ population, which represents a single statistical realisation, we find $12$ CVs in addition to the $42$ systems included in the \citet{pala20} catalog. Of those $12$, the CV with the largest ASD has a distance of $65~\rm{pc}$ and a GW frequency $f_{\rm{GW}} = 0.26~\rm{mHz}$. This system is likely an outlier, but is consistent with our adopted space density and orbital period distribution. Perhaps more important are the CVs with distances beyond $100~\rm{pc}$ that have orbital periods near $P_{\rm{min}}$, thus aiding in their GW detection while having dimmer EM signatures due to the low mass of the donor star. This population could aid in providing tighter constraints on the local space density of the CV population. Finally, we note that it is almost certainly impossible for \lisa\ to observe any orbital evolution for any of the individually resolved CVs based on the models of \citet{knigge11}. A lack of an observed orbital evolution will also limit a distance measurement or precise sky position. This is in stark contrast to AM CVn systems \citep[e.g.][]{Nelemans2004, Kremer2017, Breivik2018} and suggests that the bulk of the scientific return for the CV population will come from observations of its contribution to the \lisa\ confusion foreground. 

\begin{figure}
	\includegraphics[width=0.5\textwidth]{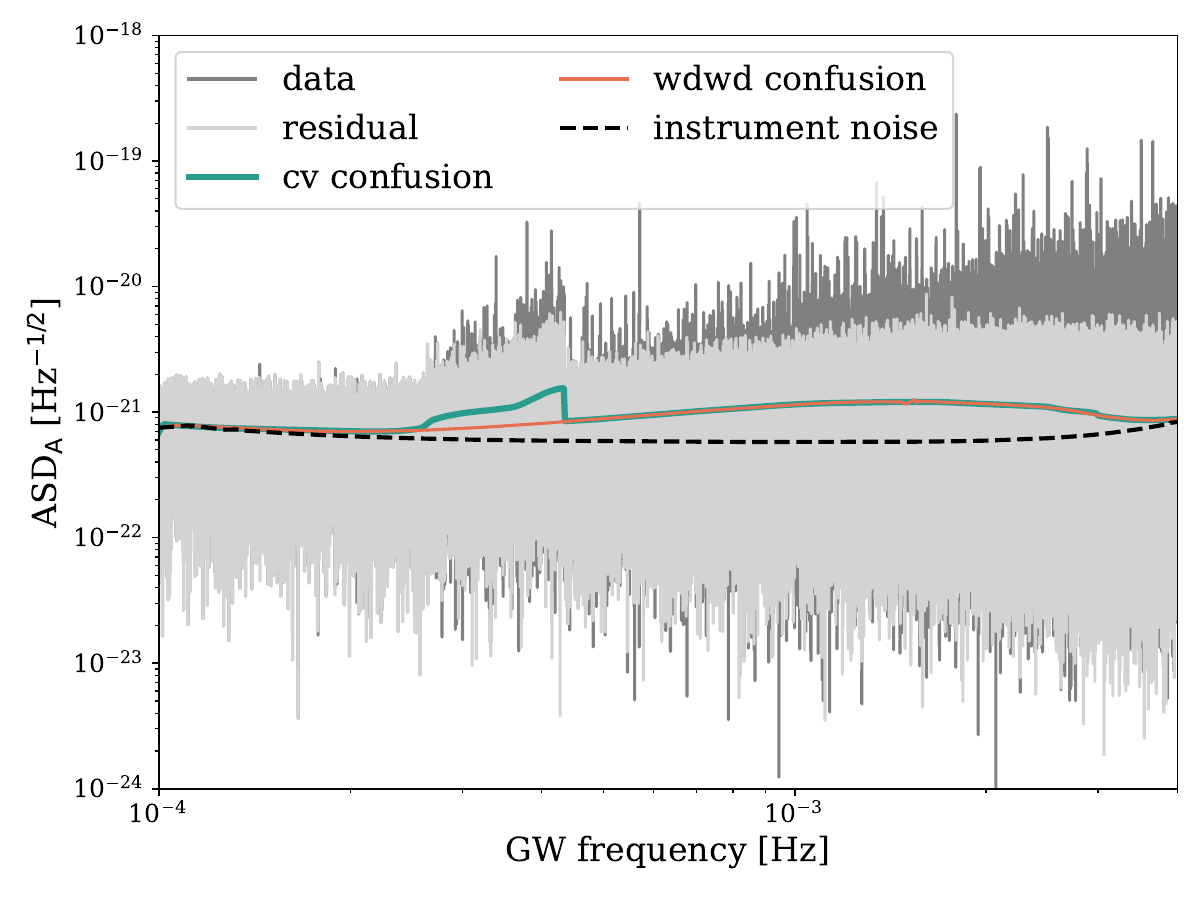}
    \caption{Amplitude spectral density of the TDI A channel (ASD$_A$)  of simulated \lisa\ data containing a galactic population of white dwarf binaries and the 1kpc sample of CVs after four years of observations. Dark gray is the original simulated data, light gray is the residual data after removal of detectable binaries, blue is the fit to the residual noise level, and orange is the fit to the residual noise level for simulations that do not include the CV population. The CVs produce a clearly identifiable bump in the residual noise spectrum and yield $\mathcal{O}(10^2)$ resolvable sources.}
    \label{fig:isd}
    \script{fig3.py}
\end{figure}

Given the high space density of CVs within $1~\rm{kpc}$ and lack of potential to observe the orbital evolution of any individual source, it is instructive to evaluate the impact of this population of the \lisa\ confusion noise. 
Fig.~\ref{fig:isd} shows the GW noise profile and confusion obtained from \texttt{ldasoft} \citep{ldasoft} for the two simulations of the Galactic population of detached DWDs discussed in Section~\ref{sec:mock_CV}. Since \texttt{ldasoft} analyzes each of the time-delay interferometry channels in the \lisa\ data stream, we consider the ASD from the A channel only. Specifically, Fig.~\ref{fig:isd} shows the ASD from the A channel of the simulated data (dark gray), the residual after all resolvable binaries have been removed from the data (light gray), the confusion noise level from the LDC population (orange) and the excess confusion noise from the CV population (green). The difference between the blue and the orange curve is the contribution to the \lisa\ foreground signal due to unresolved CVs. Interestingly we note that there are $\mathcal{O}(10^2)$ resolvable CVs in this mock population, as indicated by the difference between the dark and light gray traces at frequencies where the CV noise dominates.

\section{Discussion and Conclusions} \label{sec:discussion}

GW astronomy holds huge promise, but our ability to correctly interpret GW observations is heavily reliant on our understanding of unresolved background and foreground GW signals. This is particularly important for detecting individual GW signals in the low signal-to-noise regime. In this Letter, we consider whether CVs will contribute significantly to the GW foreground and/or the population of individually detectable GW sources for upcoming GW observatories such as \lisa\,  \textit{TianQin} (\citealt{luo16}), and the \textit{Lunar Gravitational Wave Antenna} (\citealt{harms21}).  

Focusing on \lisa\, we find that several of the closest CVs should be individually resolvable by \lisa\ within the first four years of operations. In addition, we show that the GW foreground produced by the Galactic CV population will significantly contribute to the \lisa\ mission sensitivity. More specifically, we predict a characteristic GW noise excess that rises above the currently Galactic binary populations currently included in \lisa\ sensitivity calculations in the $\approx 0.2$-$0.4$ mHz GW range. 

The excess GW noise produced by CVs is directly related their high space density and exhibits a specific form. As CVs evolve towards short orbital periods, they "pile-up" towards the so called orbital period minimum near $P_{\rm{min}} \simeq 80$~min. As a result, the GW noise excess produced by CVs increases towards the corresponding GW frequency, $f_{\rm{max}} = 2/P_{\rm{min}} \simeq 0.4$~mH, but then drops sharply for frequencies above this limit. The distinctive shape of this feature should allow its detection and characterization among other sources of GW noise. Its amplitude will then immediately allow a new and robust estimate of the Galactic CV space density, without the biases that affect all electromagnetically selected CV samples.

An important point in this context is that the CV period gap itself is largely just an electromagnetic bias. As discussed in Section~\ref{sec:intro}, the upper edge of the gap is thought to be associated with the transition of the donor star from partially radiative to fully convective. This causes a sharp reduction in the strength of MB and ultimately leads to a total loss of contact. Systems then evolve through the gap as detached binaries due to GR and perhaps residual MB until contact is re-established at the lower edge of the gap. Thus an "empty" period gap exists only in the population of {\em actively mass-transferring} CVs. The orbital period interval between $\simeq$2~hrs and $\simeq$3~hrs will contain plenty of detached WD-MS binaries, including many systems that used to be CVs (and will become CVs again). 

The orbital period distribution of {\em all} WD-MS binaries is nevertheless likely to exhibit discontinuities near the edges of the CV period gap, since the rate of period evolution is different for detached and semi-detached systems. This is the main reason we have chosen not to include detached systems in our simulations. However, since the 
evolution of detached WD-MS binaries is driven {\em solely} by GR and MB -- without the complicating effects of non-conservative mass transfer - the observed period distribution of this sub-population will provide a critical test for theoretical binary evolution models (and especially for different MB prescriptions). 

We further point out that GW observations of CVs may help in understanding the processes behind the origin of the "period gap" in CVs. This feature is thought to be a result of donor stars in CVs temporarily relaxing back to thermal equilibrium (due to becoming fully convective) and ceasing mass transfer. However, these systems will have to evolve through the period gap. \cite{knigge11} suggests that the timescale of evolution through this gap is $\approx 1$ Gyr, which in context is longer than the evolution time-scale of long-period CVs, but shorter than the evolution time-scale of systems below the gap. GW observations with \lisa\ may provide constraints on the number of systems within the period gap. This in turn could constrain the orbital period derivative of CVs within this elusive period range, and place this in the wider evolutionary context of CVs above and below the gap where the structure of the donor star is thought to be very different.

To conclude, the prospect of studying the CV population with \lisa\ and other low-frequency GW observatories is exciting for both testing conventional hypotheses on CV evolution but also to characterise and discover new CVs.

\section*{Acknowledgements}
S.S. is supported by STFC grant ST/T000244/1 and ST/X001075/1. P.J.G. is supported by NRF SARChI grant 111692. T.B.L. is supported by the NASA \lisa\ Study Office. The Flatiron Institute is supported by the Simons Foundation.

\section*{Data Availability}
This article is compiled with \texttt{showyourwork} \citep{luger21} such that the code and data to reproduce the figures in the manuscripte are used at the compilation run time. The code associated to this paper is publicly available at \href{github.com/katiebreivik/cataclysmic-pileup-pape}{github.com/katiebreivik/cataclysmic-pileup-paper} and the input data are available at \href{zenodo.org/record/7957134}{zenodo.org/record/7957134}. The scripts and datasets for each figure are linked by the Octocat and  three-stacked-cylinder icons next to each caption.



\bibliographystyle{mnras}
\bibliography{bib.bib} 






it can be placed in an Appendix which appears after the list of references.


\bsp	
\end{document}